# Multi-Affiliated Authors Behave Differently across Fields and Host Country Preferences: A Comparison in G7 and BRICS

Sichao Tong[*1,2], Liying Yang[**1]

[*] *tongsichao@mail.las.ac.cn*
ORCID: 0000-0003-0658-9963
1. National Science Library, Chinese Academy of Sciences, 100190, Beijing, China
2. Department of Library, Information and Archives Management School of Economics and Management, University of Chinese Academy of Sciences, 100190, Beijing, China

[**] *yangly@mail.las.ac.cn*
1. National Science Library, Chinese Academy of Sciences, 100190, Beijing, China

**Abstract**

This paper study author simultaneously engaged in multiple affiliations based on bibliometric data covered in the Web of Science for the 2017–2021 period. Based on the affiliation information in publication records, we propose a general classification for multiple affiliations within-country/cross-country for analyzing authors' behavior in multiple affiliations and preferences of host countries across research fields. We find a decrease in publications led by international multi-affiliated authorship after 2020, and China has shown a falling trend after 2018. More G7 countries are active in fields like Social Sciences, Clinical & Life Sciences. China, India, and Russia are active in physical sciences-related fields. Countries prefer to affiliate with G7 countries, especially in Clinical & Life Sciences. These findings may provide more insights into the understanding of the behavior and productivity of multi-affiliated researchers in the current academic landscape.

## 1. Introduction

Knowledge should be shared across organizations and disciplines to address today's societal challenges, such as climate change. Research collaboration and knowledge dissemination can be challenging, mainly when stakeholders include diverse organizations or disciplines. Can a role like a bridgebuilder be built to facilitate knowledge flow across different organizations and disciplines and remain understood throughout the flow? One of the responses is that the researchers who are being simultaneously engaged into two or more organizations (here I call these researchers multi-affiliated researchers) (European Science Foundation, 2013; Yegros-Yegros, Capponi & Frenken, 2021), who have shown a rising trend in science system in recent years (Hottenrott, Rose, & Lawson, 2021). Several studies discuss that these multi-affiliated researchers may act as bridging-person in collaborations by helping to sustain interactions between different organizations within a collaboration (European Science Foundation, 2013; Yegros-Yegros, Capponi & Frenken, 2021), as well as bring a positive effect of multiple affiliations associated with citations of publications (Hottenrott & Lawson, 2017, Huang & Chang, 2018; Sanfilippo, Hewitt, & Mackey, 2018).

Despite the potential importance of these new roles in the research landscape, they still need to be explored, especially the behavior of these researchers, and the underlying strategic determinant from different countries or regions. In light of discussions about country differences in multiple affiliations (European Science Foundation, 2013; Hottenrott, Rose & Lawson, 2021), it can be initiatives designed to improve countries' competitiveness in international research, as well as a means of mobility to facilitate research collaboration. The national needs and implicit preference reflected behind them has not yet been explored. Here we ask whether multi-affiliated researchers from different countries behave differently in different research fields and in their preference in choosing the host affiliation.

To this end, we compare multiple affiliation links of multi-affiliated authors from G7 and BRICS countries, based on their publications between 2017 and 2021, by Incites' citation topic classification system. More details can be found in the Data and Methods section. In addition, amidst the rapidly evolving landscape of the present day, characterized by events such as the prolonged COVID-19 pandemic, which emerged towards the end of 2019[1], and the China-US trade war leading to the breakdown of cooperation between the two of the largest research output producers over the world (Van Noorden, 2022), it is necessary to consider the challenges and potential strategies relevant to multi-affiliated researchers across nations. This study has been undertaken to answer the following three research questions:
- RQ1: What is the trend of multi-affiliated authors' research outputs during the COVID-19 pandemic period, and which types of authors are impacted more: those who have host affiliations within the same country, those with host affiliations abroad, or both?
- RQ2: How do multi-affiliated authors behave regarding research output across different countries, and what are the observed patterns in various disciplines?
- RQ3: Is there a preference in the choice of host affiliation among authors from different countries, and if so, how does this preference differ across countries broadly?

Through this research, we find that publications by multi-affiliated authors with multiple affiliations within-country/cross-country have shown different trends during the period of COVID-19 period, especially Chinese multi-affiliated authors have shown a relatively sharp decrease after 2018, the year of China-USA conflict erupted, which can also be seen in Van Noorden's study (2022). Multi-affiliated authors from G7 and BRICS countries show different activity across research fields. Moreover, by defining the home country and host country for such authors in multiple affiliation links, we find that authors from G7 and BRICS countries show different preferences in choosing their host-affiliated countries. Our study enriched explorations and explanations of multiple affiliations' underlying strategical needs and implicit preferences. These findings might further research to better understand the role of multi-affiliated researchers.

**2. Data and Methods**

*2.1. Data collection*
We retrieved Web of Science (WoS) Core Collection (SCI, SSCI) publications published between 2017 and 2021 with all author address records. Only publications of the Web of Science document types "Article" and "Review" are included in the data collection. Manual institutional disambiguation was undertaken.

The Incites citation topic classification system is used here to see the different multiple affiliations patterns across research fields, as listed in Table 1. We utilize the macro-level topic to see the statistics and patterns from a coarse view, resulting in 10 broader science fields. We here remove the field of Art and humanities, for the insufficient coverage of WoS datasets. Table 1 also lists the number of publications by field we used.

Table 1. Number and share of publications, by research field.

---
[1] https://en.wikipedia.org/wiki/COVID-19

| Field | Field Label (Abbreviation) | N of Publications | Share (%) |
|---|---|---|---|
| Social Sciences | Soc. Sci. | 663,756 | 8% |
| Clinical & Life Sciences | Cli. & Life Sci. | 3,122,743 | 37.5% |
| Earth Sciences | Earth Sci. | 325,788 | 3.9% |
| Agriculture, Environment & Ecology | Agr. & Env. & Eco. | 965,929 | 11.6% |
| Physics | Physics | 551,795 | 6.6% |
| Chemistry | Chemistry | 1,222,190 | 14.7% |
| Engineering & Materials Science | Eng. & Mater. Sci. | 585,689 | 7% |
| Electrical Engineering, Electronics & Computer Science | EE & Elec. & CS | 659,393 | 7.9% |
| Mathematics | Mathematics | 219,307 | 2.6% |

Note: here, fields are ordered by first social sciences-related research field, secondly biomedical-related, then physical sciences-related, and finally engineering and mathematics-related.

Countries from the Group of G7 (Canada, France, Germany, Italy, Japan, the United Kingdom, and the USA)[2] and BRICS (Brazil, China, India, Russia, and South Africa)[3] are used to compare the differences between multiple affiliations patterns (Table 2).

Table 2. Number of first affiliated publications, by country.

| Country Group | Country Name | N of Publications (first affiated) |
|---|---|---|
| G7 | USA | 1,530,208 |
| | UK | 365,292 |
| | Germany | 357,934 |
| | Japan | 320,269 |
| | Italy | 261,577 |
| | Canada | 228,191 |
| | France | 218,870 |
| BRICS | China | 1,962,012 |
| | India | 312,962 |
| | Brazil | 208,112 |
| | Russia | 139,520 |
| | South Africa | 43,434 |

Note: here, we present the number of publications for the first affiliated country (first/corresponding author only). Results have been listed in descending order by country group and publication volumes.

*2.2. Classification of multiple affiliations, multi-affiliated authorship and publications*
There are two types of multi-affiliated authorship in our dataset:

➢ A_NM: the national multi-affiliated author who is affiliated with two or more institutions from one country.
➢ A_IM: the international multi-affiliated author who is affiliated with two or more institutions from two or more countries.

---

[2] https://en.wikipedia.org/wiki/G7
[3] https://en.wikipedia.org/wiki/BRICS

For a publication, we divide the authorship of a publication into two types: main and ordinary contributor, and we consider the first author and corresponding author as the main author for a publication. Here we focus on the main authorship in our study.

Given this, we also propose a classification to define multiple affiliations at the level of publication, to analyze their scientific output:

- ➤ P_NM: the publication with the national multi-affiliated main author, and without an international multi-affiliated main author.
- ➤ P_IM: the publication with the international multi-affiliated main author.

Thus, the combination of multiple affiliated organizations, i.e., multiple affiliation links, happens naturally via multi-affiliated authors as follows:

- ➤ NM: multiple affiliations of two or more institutions from one country.
- ➤ IM: multiple affiliations of institutions from two or more countries.

*2.3. Indicators*

*2.3.1 Activity index*
Frame (1977) introduced the activity index to characterize the relative research effort a country devotes to a given research field. It's also similar to revealed competitive advantage (RCA) index proposed by Balassa (1965). With respect to a research field, the activity index (AI) of a country during the given period can be defined as:

$$AI_{c,f} = \frac{P_{c,f} / \sum_{i=1}^{n} P_{c,f}}{P_{w,f} / \sum_{i=1}^{n} P_{w,f}}$$

Where $P_{c,f}$ is the publications of country $c$ in field $f$, $\sum_{i=0}^{n} P_{c,f}$ is the total publications of country $c$ from all fields, $P_{w,f}$ is the publications of the world in field $f$, $\sum_{i=0}^{n} P_{w,f}$ is the total publications of country the world from all fields.

Since here this study look at the output led by multi-affiliated authors, we can define AI index correspondingly:

$$AI_{c,f,NM} = \frac{P_{c,f,NM} / \sum_{i=1}^{n} P_{c,f,NM}}{P_{w,f,NM} / \sum_{i=1}^{n} P_{w,f,NM}}$$

$$AI_{c,f,IM} = \frac{P_{c,f,IM} / \sum_{i=1}^{n} P_{c,f,IM}}{P_{w,f,IM} / \sum_{i=1}^{n} P_{w,f,IM}}$$

*2.3.2 Gini coefficient*
Developed by Corrado Gini as a measure of income inequality, the Gini coefficient, defined as the mean of differences between all pairs of individuals (David, 1968), has been used to address various topics. There is a theoretical maximum of 1 and a theoretical minimum of 0. Here with respect to host country-share inequality, the Gini coefficient of a country in building multiple affiliation links with other countries can be defined as[4]:

$$G_c = \frac{\sum_{i=1}^{n}(2i - n - 1)x_i}{n \sum_{i=1}^{n} x_i}$$

Where $x_i$ represents the value of multiple affiliation links build with each host country, $n$ represents the total number of host countries of observed country, and $i$ represents the rank of $x_i$.

*2.4. International multi-affiliated authorship: home country and host country*
Our methodological approach posits that the first affiliation listed represents the primary affiliation of the multi-affiliated author. Consequently, for A_IM, following (Hottenrott, Rose, & Lawson, 2021), we designate:

➢ Home country: the first country of affiliations of the multi-affiliated author
➢ Host country: any other country of affiliations of the multi-affiliated author.

## 3. Results

*3.1. Trends in publications led by multi-affiliated authorship*
Publications with the national multi-affiliated authorship are more evident in Figure 1, where about 17.5% of scientific output are led by the national multi-affiliated authorship in 2021, while publications led by international multi-affiliated authorship account for 6%. We also observe a decrease in publications led by international multi-affiliated authorship after 2020. Issues such as Covid-19 may play a role here.

Figure 1: Share of publications in all scientific output by the type of multi-affiliated authorship: national multi-affiliated authorship, international multi-affiliated authorship.

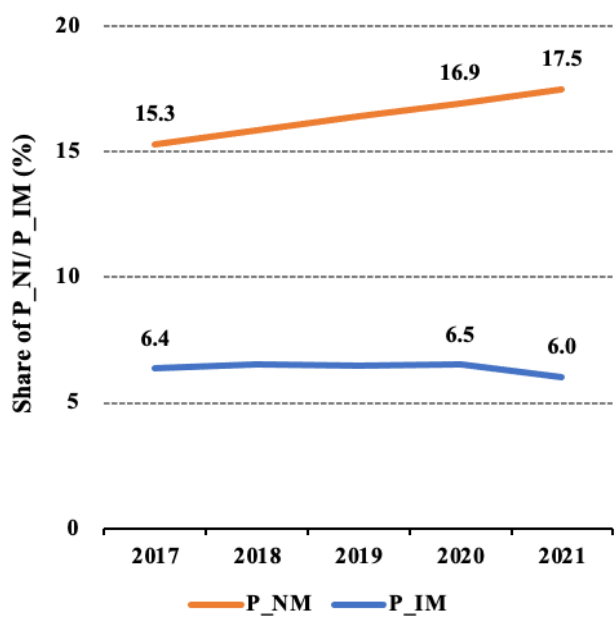

---
[4] https://www.statsdirect.com/help/default.htm#nonparametric_methods/gini.htm?TocPath=Nonparametric%2520methods%257C_____19

When we go further into the type of publications led by international multi-affiliated authorship, trends during the Covid-19 periods are various when considering different countries. Figure 2 shows that all G7 countries show a decrease. In contrast, three of the BRICS countries-South Africa, Brazil, and Russia, show an increase in the type of publications led by the international multi-affiliated authorship. In addition, China has shown a falling trend after 2018, and issues such as the China-USA conflict may play a role here.

Figure 2: Share of publications led by international multi-affiliated authorship in all scientific output by country: G7 and BRICS

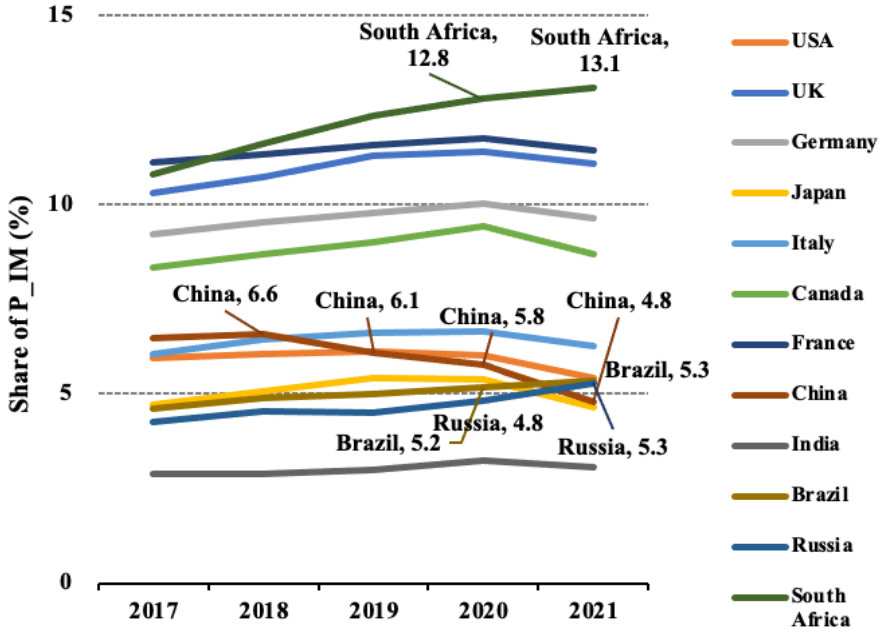

*3.2. Field-based activity of multi-affiliated authorship*
This section calculated the activity index for each combination of 9 fields and G7 and BRICS countries (Figure 3). Here we will not provide an in-depth analysis of the phenomenon of multiple affiliations, but rather a glimpse of how geography, culture, economics, and science might influence the behavior of different countries regarding multiple affiliations.

Figure 3: Activity of multi-affiliated authorship per country broken down by research field classification. a) national multi-affiliated authorship, b) international multi-affiliated authorship. Color represents the value of the activity index.

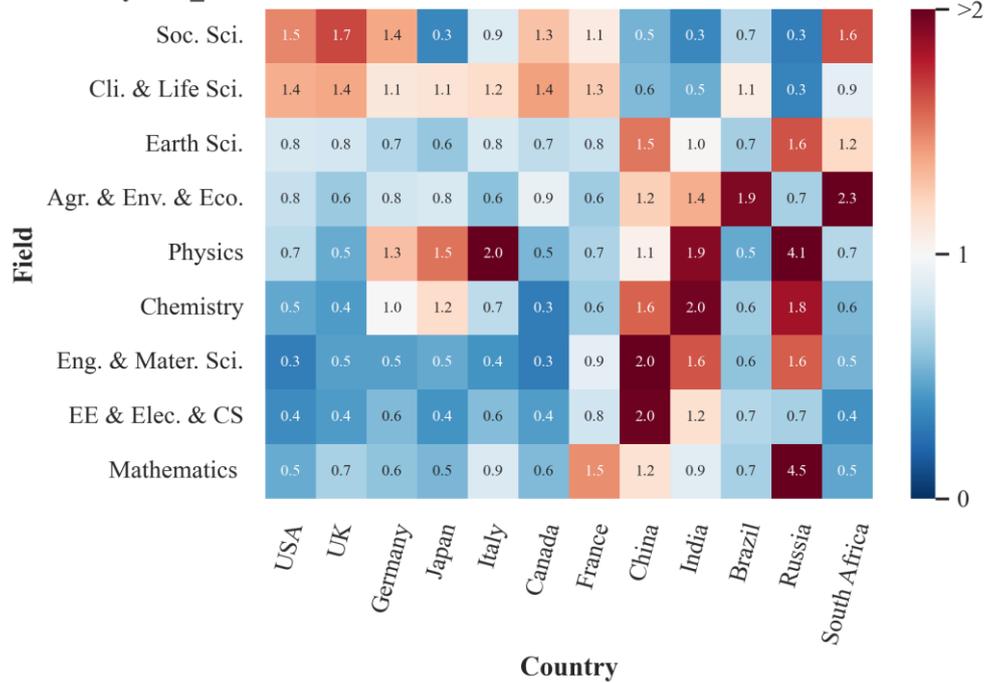

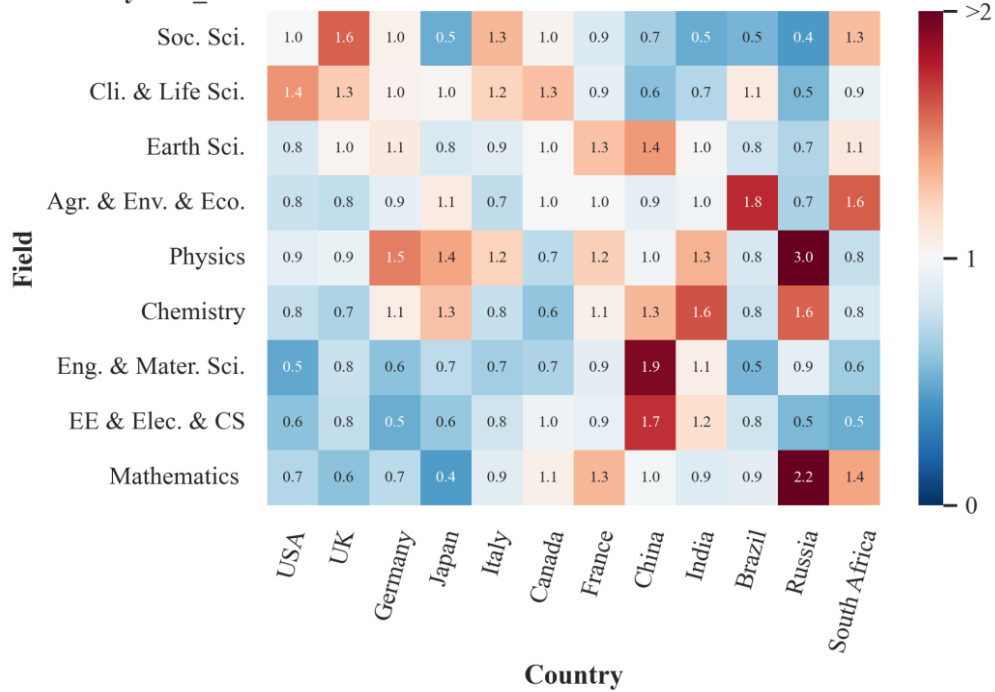

For instance, Figure 3a and Figure 3b both show that G7 countries are more active in Clinical & Life Sciences, which has been potentially associated with the substantial investment capacity and resources in clinical and biomedical research in developed countries (Miao, Murray, Jung, Larivière, Sugimoto, & Ahn, 2022). This phenomenon can also be seen in Social Sciences. Japan in the G7 group is more active in Physics and Chemistry. In BRICS countries, China and India perform more in chemistry, EE & Elec fields. & CS and Eng. & Mater. Sci. Russia active more in Physics and Mathematics. Brazil and South Africa are more active in the field of Agr. & Env. & Eco.

*3.3. Inequality combinations in host affiliated countries.*

Next, we emphasize the potential preference regarding building links with host countries. In this case, we analyze the dataset among international multi-affiliated authorship based on the definition of home country and host country.

Figure 4: Gini coefficient of host country-share inequality per country in building multiple affiliation links with other countries.

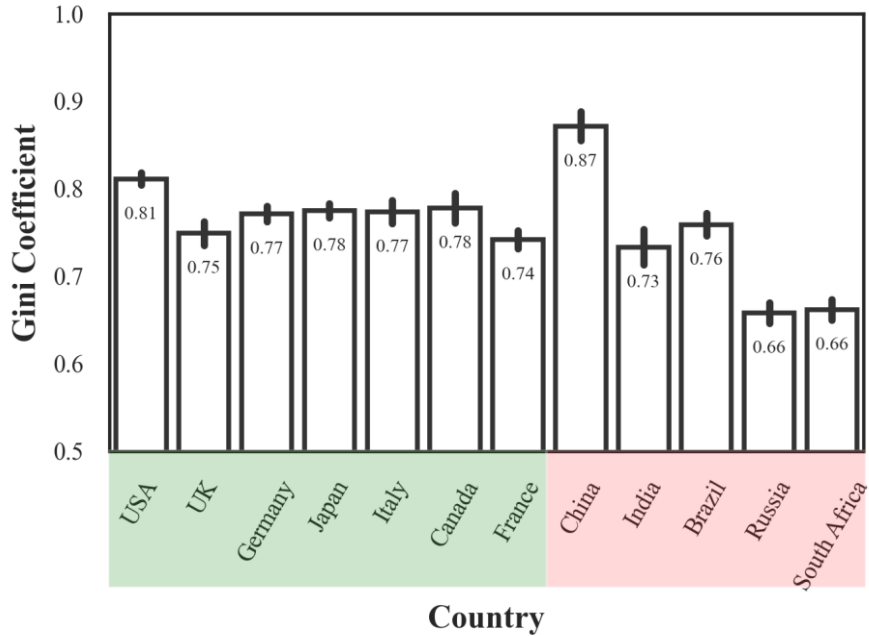

As shown in Figure 4, the most-affiliated host countries garner higher shares of multiple affiliation links in these countries. All gini coefficients of host country-share are higher than 0.6, and BRICS countries like India, Russia and South Africa have lower gini coefficients. China has a most significant Gini coefficient of affiliated host country-share of around 0.87, which means it concentrates on building multiple affiliation links in a small share of its most affiliated host countries.

*3.4. Preference of host country*

Next, we combine these two country groups (i.e., G7 countries and BRICS countries) to investigate their preference of the host country group per research field. Figure. 5 shows the difference between G7 and BRICS and between different research fields.

Figure 5: Preference of host country group per research field. a) Preference of building multiple affiliation links with G7, b) Preference of building multiple affiliation links with BRICS. Global baselines are shown.

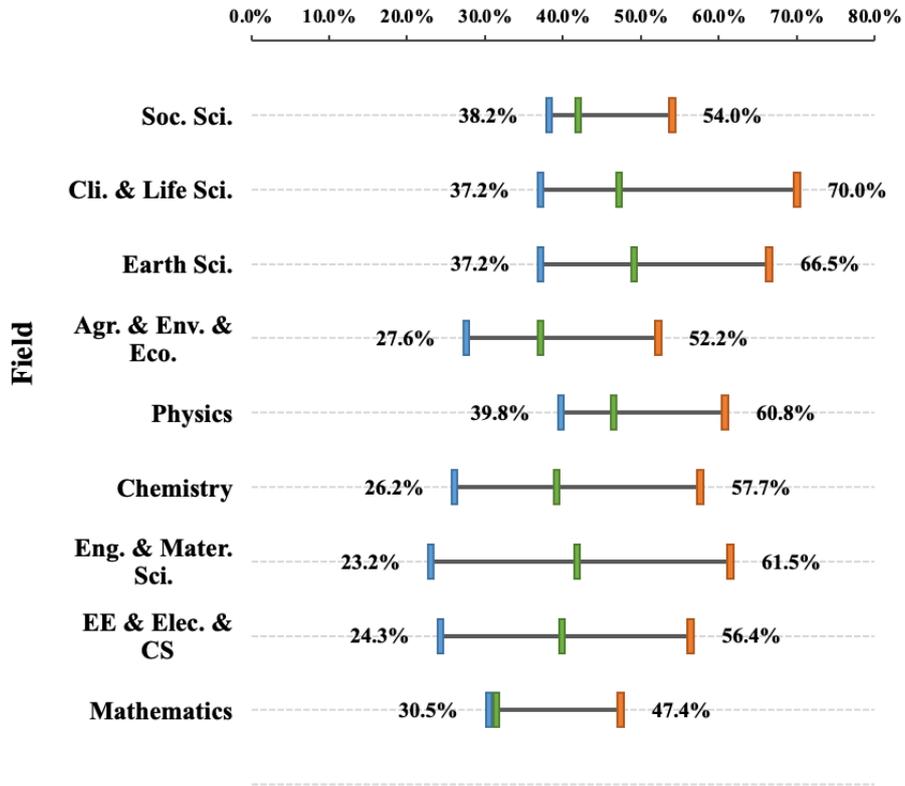

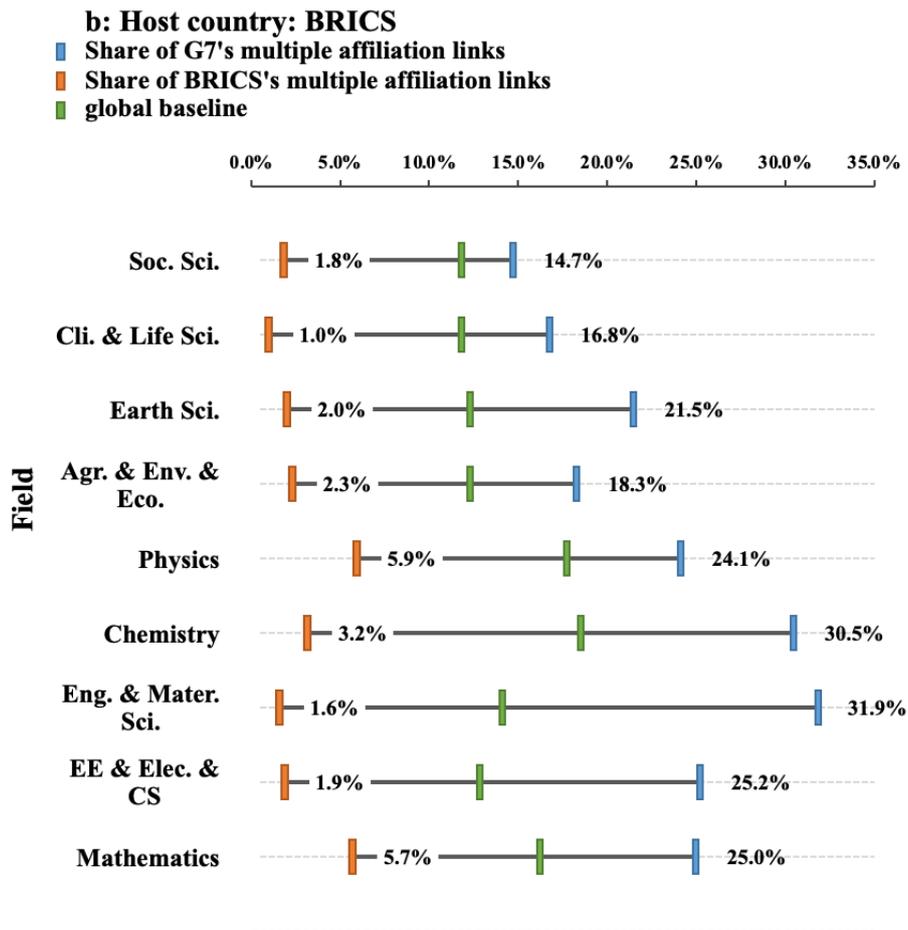

Figure. 5 visualizes the difference between G7 and BRICS in 'prefer to affiliate G7'/ 'prefer to affiliate BRICS'. As expected, in all fields, the behavior of 'prefer to affiliate G7' is more prominent. We further investigate that compared to G7 countries, BRICS show a more preference to affiliate G7, and vice versa, i.e., compared to BRICS, G7 prefers more to affiliate BRICS. Moreover, BRICS shows a significantly high preference to affiliate G7 in Clinical & Life Sciences (70% of multiple affiliation combinations happen with G7). In contrast, G7 shows a higher preference to affiliate BRICS in Chemistry and Eng. & Mater. Sci., with more than 30% of their multiple affiliation combinations, happen with BRICS.

## 4. Discussion

Through an exploration of the author who simultaneously engaged in multiple affiliations based on bibliometric data, we try to answer the questions presented previously in the introduction section, mainly focusing on the trend of scientific output led by multi-affiliated authorship, and activity of multi-affiliated researchers in various research fields, by different countries, as well as the preference of affiliate with the host country.

These findings suggest that crises like Covid-19, geopolitical tensions, and underlying strategic determinants from different countries or regions may affect multi-affiliated authors' behaviors, especially in building multiple affiliation links in different research fields and preferences in forming affiliation links with other countries.

Nevertheless, this study currently has several limitations. Discipline schema is one of them. We use Incites dataset without enough coverage of data from Art and humanities. Another is that we only analyze G7 and BRICS countries in this study. The multi-affiliated authorships of more countries with different strategic or economic status needs further investigation. Moreover, based on current results, we need a further in-depth explanation of the multiple affiliations phenomenon.

The findings would contribute to a better understanding of the behavior and productivity of multi-affiliated researchers in the current academic landscape, highlighting the impact of the crisis, research priorities, and collaboration patterns.

**Acknowledgments**
The authors thank Ting Yue, Zhesi Shen, and Li Li for helpful discussions, and Fuyou Chen, Jiandong Zhang for help in data processing.

**Author contributions**
Conceptualization: Sichao Tong, Liying Yang
Data curation: Sichao Tong
Formal analysis: Sichao Tong
Investigation: Sichao Tong
Methodology: Sichao Tong, Liying Yang
Visualization: Sichao Tong
Writing – original draft: Sichao Tong
Writing – review & editing: Sichao Tong, Liying Yang
Supervision: Liying Yang

**Competing interests**
Authors have no competing interests.